\documentclass[english]{revtex4}
\usepackage[T1]{fontenc}
\usepackage[latin9]{inputenc}
\usepackage{amssymb}
\usepackage{esint}
\usepackage{babel}

\begin{document}

\title{Thermodynamics of the $\left(2+1\right)$-dimensional Black Hole
with non linear Electrodynamics and without Cosmlogical Constant from the Generalized Uncertainty Principle}

\author{Alexis Larrañaga}

\affiliation{Universidad Nacional de Colombia. Observatorio Astronómico Nacional
(OAN)}

\affiliation{Universidad Distrital Francisco José de Caldas. Facultad de Ingeniería.
Proyecto Curricular de Ingeniería Electrónica}

\email{ealarranaga@unal.edu.co}

\begin{abstract}
In this paper, we study the thermodynamical properties of the $\left(2+1\right)$
dimensional black hole with a non-linear electrodynamics and without
cosmological constant using the Generalized Uncertainty Principle
(GUP). This approach shows that there is a maximum temperature for
the black hole depending only on the electric charge and corresponding to the minimum radius of the event horizon, of the order of the Planck scale. Finally, we show that the heat capacity
for this black hole has the expected behavior.
\end{abstract}
\maketitle

In $\left(2+1\right)$ dimensional gravity, the charged black hole
(static BTZ solution) has an electric field that is proportional to
the inverse of $r$, hence its potential is logarithmic. If the source
of the Einstein equations is the stress-energy tensor of non-linear
electrodynamics, which satisfies the weak energy conditions, one can
find a solution with a Coulomb-like electric field (proportional to
the inverse of $r^{2}$). This kind of solution was reported by Cataldo
et. al. \cite{cataldo}, and describes charged-AdS space when considering
a negative cosmological constant. 

The thermodynamical properties of black holes are associated with
the presence of the event horizon. As shown recently \cite{larranagagarcia},
the threedimensional black hole with nonlinear electrodynamics satisfies
a differential first law with the usual form 

\begin{equation}
dM=TdS+\Phi dQ,\end{equation}
where $T$ is the Hawking temperature that can be expressed in terms
of the surface gravity at the horizon $\kappa$ by

\begin{equation}
T=\frac{\kappa}{2\pi}.\label{eq:hawkingtemp}\end{equation}

However, in recent years the uncertainty relation that includes gravity
effects, known as the Generalized Uncertaintly Principle (GUP), has
shown interesting results in the context of black hole evaporation
\cite{arraut}, extending the relation between temperature and mass
to scales of the order of the Planck lenght, $l_{p}=1.61\times10^{-33}cm$.
This treatment has shown that $l_{p}$ is the smallest length scale
in the theory and it is related to the existence of an extreme mass
(the Planck mass $m_{p}=1.22\times10^{19}GeV$, which becomes the
black hole remnant), that corresponds to the maximum possible temperature. 

In this paper we investigate the thermodynamics of the three-dimensional
black hole with a nonlinear electric field reported in \cite{cataldo},
to show how the GUP can be used to calculate the Hawking temperature
associated with the black hole. The $T\left(M\right)$ equation gives
the usual relation for small masses but gets deformed when the mass
becomes larger. We also show how there is a maximum temperature for
the black hole that corresponds to a horizon with size in the Planck
scale. Finally we calculate the heat capacity for this black hole,
to show that it has the right physical behavior.

\section{The 3-dimensional Black Hole with non-linear Electrodynamics}

The metric reported by Cataldo et. al. \cite{cataldo} is a solution
of the $\left(2+1\right)$ dimensional Einstein's field equations
with a negative cosmological constant $\Lambda<0$,

\begin{equation}
G_{\mu\nu}+\Lambda g_{\mu\nu}=8\pi GT_{\mu\nu}.\end{equation}
To obtain a Coulomb-like electric field, Cataldo et. al. used a nonlinear
electodynamics. In the non-linear theory, the electromagnetic action
$I$ does not depend only on the invariant $F=\frac{1}{4}F_{\mu\nu}F^{\mu\nu}$,
but it can be a generalization of it, for example

\begin{equation}
I\propto\int d^{3}x\sqrt{\left|g\right|}\left(F_{\mu\nu}F^{\mu\nu}\right)^{p},\end{equation}
where $p$ is some constant exponent. If the energy-momentum tensor
is restricted to be traceless, the action becomes a function of $F^{3/4}$,
and the static circularly symmetric solution obtained has the line
element 

\begin{equation}
ds^{2}=-f\left(r\right)dt^{2}+\frac{dr^{2}}{f\left(r\right)}+r^{2}d\varphi^{2},\label{eq:staticmetric}\end{equation}
where

\begin{equation}
f\left(r\right)=-M-\Lambda r^{2}+\frac{4GQ^{2}}{3r}.\end{equation}

The electric field for this solution is

\begin{equation}
E\left(r\right)=\frac{Q}{r^{2}},\label{eq:electric field}\end{equation}

which is the standard Coulomb field for a point charge. The metric
depends on two parameters $Q$ and $M$, that are identified as the
electric charge and the mass, respectively. The horizons of this solution
are defiened by the condition \begin{equation}
f\left(r\right)=0\end{equation}
 or

\begin{equation}
-M-\Lambda r^{2}+\frac{4GQ^{2}}{3r}=0.\label{eq:horizonmass}\end{equation}

However, if we consider a zero cosmological constant, $\Lambda=0$,
the resulting black hole has interesting properties. The line element
becomes 

\begin{equation}
ds^{2}=-\left(-M+\frac{4GQ^{2}}{3r}\right)dt^{2}+\frac{dr^{2}}{\left(-M+\frac{4GQ^{2}}{3r}\right)}+r^{2}d\varphi^{2},\end{equation}
that shows how this spacetime is asymptotically flat. Note that this
charged black hole has just one horizon, located at

\begin{equation}
r_{H}=\frac{4GQ^{2}}{3M}.\label{eq:horizon}\end{equation}

The Hawking temperature for this black hole is given by the usual
definition (\ref{eq:hawkingtemp}), where the surface gravity can
be calculated as

\begin{equation}
\kappa=\chi\left(x^{\mu}\right)a,\end{equation}
with $a$, the magnitude of the four-acceleration and $\chi$, the
red-shift factor. In order to calculate $\chi$, we will consider
a static observer, for whom the red-shift factor is just the proportionality
factor between the timelike Killing vector $K^{\mu}$ and the four-velocity
$V^{\mu}$, i.e.

\begin{equation}
K^{\mu}=\chi V^{\mu}.\end{equation}
 The metric (\ref{eq:staticmetric}) has the Killing vector

\begin{equation}
K^{\mu}=\left(1,0,0\right)\end{equation}
while the four-velocity is calculated as

\begin{equation}
V^{\mu}=\frac{dx^{\mu}}{d\tau}=\left(\frac{dt}{d\tau},0,0\right).\end{equation}
This gives

\begin{equation}
V^{\mu}=\left(f^{-1}\left(r\right),0,0\right)=\left(\frac{1}{\sqrt{-M+\frac{4GQ^{2}}{3r}}},0,0\right),\end{equation}
and therefore, the red-shift factor is

\begin{equation}
\chi\left(r\right)=\sqrt{-M+\frac{4GQ^{2}}{3r}}.\end{equation}
 On the other hand, the four-acceleration is given by

\begin{equation}
a^{\mu}=\frac{dV^{\mu}}{d\tau},\end{equation}
that has components

\begin{eqnarray}
a^{0}=a^{\varphi} & = & 0\\
a^{r} & = & -\frac{2}{3}\frac{GQ^{2}}{r^{2}}.\end{eqnarray}
and therefore, the magnitude of the four-acceleration is

\begin{equation}
a=\sqrt{g_{\mu\nu}a^{\mu}a^{\nu}}=\frac{-\frac{2}{3}\frac{GQ^{2}}{r^{2}}}{\sqrt{-M+\frac{4GQ^{2}}{3r}}}.\end{equation}

Then, the surface gravity at the event horizon is given by the absolute
value

\begin{equation}
\kappa=\left|-\frac{2}{3}\frac{GQ^{2}}{r^{2}}\right|_{r=r_{H}}\end{equation}

\begin{equation}
\kappa=\frac{2GQ^{2}}{3r_{H}^{2}}=\frac{3M^{2}}{8GQ^{2}}\end{equation}

and the Hawking temperature is,

\begin{equation}
T=\frac{GQ^{2}}{3\pi r_{H}^{2}}=\frac{3M^{2}}{16\pi GQ^{2}}.\label{eq:bhtemperature}\end{equation}

\section{Hawking Radiation and the Generalized Uncertainty Principle}

Now we will consider the GUP and the Hawking radiation derived from
it. We will show how the GUP will produce a deformation in the temperature-mass
relation when considered close to the Planck lenght. As stated by
Adler and Santiago \cite{adler} the GUP is given, in units with
$c=1$, by the relation

\begin{equation}
\Delta x\Delta p\gtrsim\hbar+G\hbar\left(\Delta p\right)^{2}.\end{equation}
where $G$ is the gravitational constant and $\hbar$ is the Planck
constant. Since the Planck lenght $l_{p}$ can be written as

\begin{equation}
l_{p}^{2}=G\hbar=\frac{\hbar^{2}}{m_{p}^{2}},\end{equation}
where $m_{p}$ is the mass of Planck, the GUP can be written as

\begin{equation}
\Delta x\Delta p\gtrsim1+l_{p}^{2}\left(\Delta p\right)^{2},\end{equation}
or

\begin{equation}
\Delta x\Delta p\gtrsim1+\frac{\left(\Delta p\right)^{2}}{m_{p}^{2}},\end{equation}
in units with $\hbar=1$. To apply this uncertainty principle to the
black hole evaporation process consist in identifying the $\Delta x$
with the event horizon radius $r_{H}$ and the momentum $\Delta p$
with the Hawking temperature up to a $2\pi$ factor \cite{myung}.
Therefore, we can write the GUP as a quadratic equation for the temperature,

\begin{equation}
r_{H}2\pi T=1+\frac{4\pi^{2}T^{2}}{m_{p}^{2}}\end{equation}

\begin{equation}
T^{2}-\frac{2Q^{2}}{3\pi M}T+\frac{m_{p}^{2}}{4\pi^{2}}=0.\end{equation}
From which it follows that the temperature-mass relation is

\begin{equation}
T\left(M\right)=\frac{Q^{2}}{3\pi M}\left[1+\sqrt{1-\frac{9M^{2}m_{p}^{2}}{4Q^{4}}}\right],\label{eq:bhtemperature2}\end{equation}
where we have chosen the plus sign of the root in order to obtain the
black hole temperature (\ref{eq:bhtemperature}) in the limit of small
$M$. Note that the argument in the square root defines a maximum
mass for the black hole,

\begin{equation}
M^{max}=\frac{2Q^{2}}{3m_{p}}.\end{equation}
This mass gives, using equation (\ref{eq:bhtemperature}), the maximum
temperature permited to the black hole,

\begin{equation}
T^{max}=\frac{Q^{2}}{12\pi}.\end{equation}
Equation (\ref{eq:horizon}) implies that the black hole with the
maximum temperature has an event horizon with a minimum radius

\begin{equation}
r_{H}^{min}=\frac{2}{m_{p}}=2l_{p}.\end{equation}
Therefore, we conclude that $r_{H}$ can not be smaller than twice
the Planck lenght. On the other hand, the heat capacity for the $\Lambda=0$
black hole can be calculated using equation (\ref{eq:bhtemperature2}), 

\begin{eqnarray}
C_{Q} & = & \left(\frac{\partial M}{\partial T}\right)_{Q}\\
 & = & -\frac{3\pi M}{Q^{2}}\left[\frac{1}{M}\left(1+\sqrt{1-\frac{9M^{2}m_{p}^{2}}{4Q^{4}}}\right)+\frac{9M^{2}m_{p}^{2}}{4Q^{4}}\left(1-\frac{9M^{2}m_{p}^{2}}{4Q^{4}}\right)^{-1}\right]^{-1}.\end{eqnarray}
Since the heat capacity of this black hole is always negative, $\frac{\partial T}{\partial M}<0$,
the behavior of $T$ is the expected, as $M$ increases, the temperature
decreaces\@.

\section{Conclusion}

We have studied the thermodynamics of the $\left(2+1\right)$ dimensional
black hole with non-linear electrodynamics and without cosmological
constant using the Generalized Uncertainty Principle. This gives a
maximum mass for the black hole, that corresponds to a maximum Hawking
temperature depending only on the electric charge $Q$. The solution
with the maximum temperature is obtained when the black hole has a
size of the order of the Planck scale (minimum horizon). 

Equation (\ref{eq:bhtemperature2}) gives the temperature-mass relation,
and as is shown, it gives the standard Hawking temperature (\ref{eq:bhtemperature})
for small masses, but gets deformed for masses close to $M^{max}$.
Finally, the heat capacity of this black hole is negative, giving
the right physical behavior.

This analysis confirms that Planck lenght seems to be the smallest
lenght in nature, even in $\left(2+1\right)$ dimensions. In a forthcoming
paper, consequences of the GUP in the $\left(2+1\right)$ dimensional
black hole with non-lineal elctrodynamics and non-zero cosmological
constant will be discussed.

\end{document}